\documentclass[12pt]{iopart}

\begin{document}

\title{Szekeres models: a covariant approach}
\author{Pantelis S. Apostolopoulos$^1$}

\address{$^1$Technological Educational Institute of Ionian Islands, Department
of Environmental Technology, Island of Zakynthos, Greece}

\ead{papost@teiion.gr; papost@phys.uoa.gr \\
\textbf{PACS}: 04.20.-q, 4.20.Nr, 98.80.Jk}

\begin{abstract}
We exploit the 1+1+2 formalism to covariantly describe the inhomogeneous and
anisotropic Szekeres models. It is shown that an \emph{average scale length}
can be defined \emph{covariantly} which satisfies a 2d equation of motion
driven from the \emph{effective gravitational mass} (EGM) contained in the
dust cloud. The contributions to the EGM are encoded to the energy density
of the dust fluid and the free gravitational field $E_{ab}$. We show that
the quasi-symmetric property of the Szekeres models is justified through the
existence of 3 independent \emph{Intrinsic Killing Vector Fields (IKVFs)}.
In addition the notions of the Apparent and Absolute Apparent Horizons are
briefly discussed and we give an alternative gauge-invariant form to define
them in terms of the kinematical variables of the spacelike congruences. We
argue that the proposed program can be used in order to express Sachs'
optical equations in a covariant form and analyze the confrontation of a
spatially inhomogeneous irrotational overdense fluid model with the
observational data.
\end{abstract}

\maketitle

\section{Introduction}

\setcounter{equation}{0} The discovery of the accelerated phase of the
Universe \cite{SuperNovaTeam1} and the detailed chartography of the
temperature anisotropies in the Cosmic Microwave Background (CMB) \cite%
{accel2, wmap, Ade:2013zuv} emerge the fact that additional degrees of
freedom are necessary in order to fit the observational data to the
underlying geometry. The current observational \textquotedblleft
strategy\textquotedblright\ states that the Universe is \textquotedblleft
almost\textquotedblright\ homogeneous and isotropic at very large scales ($%
\sim $ 100 Mpc) which imposes the highly symmetric
Friedmann-Lemaitre-Robertson-Walker (FLRW) geometry as the standard
cosmological model. Within the FLRW spacetime, the accelerated expansion is
usually interpreted by introducing a $\Lambda -$term dark sector which
affect the cosmological evolution whereas the small temperature fluctuations
occur due to the presence of local inhomogeneities and are described as
first order perturbations of the background model.

However one can then argue that the presence, the form and the evolution of
local density and expansion inhomogeneities and anisotropies seeded the
formation in the Universe at small or medium scales ($<$ 10 Mpc), influence
the travel of light rays causing various effects like focusing, lensing thus
giving an alternative explanation of the SN data out of the $\Lambda $CDM
scheme. Therefore one must take into account exact inhomogeneous models
which can be seen as \textquotedblleft perturbations\textquotedblright\ in a
FLRW background. In this respect there is an increased interest of using 
\emph{inhomogeneous} configurations within the standard cosmological
paradigm like the Lemaitre-Tolman (LT) model or the Szekeres models which
represent an immediate generalisation of the former (see \cite%
{Bolejko:2011jc} and papers cited therein). Both can be matched to the FLRW
at large scales and it has been shown that they confront with the
observational data without the need of a cosmological constant, due to the
rich variety of the matter density profiles accommodating these models.

At first glance, Szekeres solution \cite{Szekeres:1974ct, Krasinski} (or its
generalization to include a non-zero cosmological constant \cite%
{Barrow-Stein-Schabes}) seems to be general enough within the class of
inhomogeneous models due to the non-existence of isometries i.e. Killing
Vector Fields (KVFs) \cite{Bonnor1, Bonnor-Sulaiman-Tomimura} or, as far as
we know, any other kind of symmetry. Nevertheless these models exhibit
various special features (like the conformal flatness of the hypersurfaces $%
t=$const. \cite{Berger:1977qh}) which make them more tractable than expected 
\cite{Apostolopoulos:2006zn}. In this paper we provide a covariant way to
reproduce the key ingredients and describe, in a gauge-invariant form, the
family of Szekeres models by using the well established theory of spacelike
congruences. In fact the 1+1+2 splitting technique is the most appropriate
tool for analyzing the Szekeres models due to the decouple of the spatial
divergence and curl equations from the evolution equations of the
kinematical and dynamical variables and the existence of 2d hypersurfaces of
constant curvature at $t=$const and $r=$const. As a result the formalism
presented here can be used, in principle, to analyze light propagation and
structure formation within a FLRW background.

The paper is organized as follows: in section 2 we give the evolution and
constraint (spatial derivatives) equations of the associated kinematical and
dynamical quantities. A further 1+2 splitting is applied in section 3,
revealing the role of the corresponding kinematical variables of the
spacelike congruences in the dynamics. Using these results, in section 4, we
define the \emph{rate of change or the expansion rate of the 2d surface area}
and the \emph{effective gravitational mass} of the shells of dust which is
constituted of the total energy density $\rho $ of the matter fluid and the
contribution from the free gravitational field (encoded in the electric part
of the Weyl tensor). For the quasi-spherical case, due to the constant
curvature feature of the hypersurface $t,r=$constant, we show that there
exists a $SO(2)$ group of \emph{Intrinsic Killing Vector Fields} (IKVFs). We
conclude in section 5 and discuss briefly the notions of Apparent Horizons
(AH) and Absolute Apparent Horizons (AAH) of the Szekeres models \cite%
{Hellaby:2002nx} by proposing a covariant definition of them. We also
discuss possible application of the suggested program regarding the
influence of the gravitational field on the light beams that come from a 
\emph{specific direction} of an isotropic source (S) and passing through
certain fluid configurations.

Throughout this paper, the following conventions have been used: the pair ($%
\mathcal{M},\mathbf{g}$) denotes the spacetime manifold endowed with a
Lorentzian metric of signature ($-,+,+,+$), spacetime indices are denoted by
lower case Latin letters $a,b,...=0,1,2,3$, spacelike eigenvalue indices are
denoted by lower case Greek letters $\alpha ,\beta ,...=1,2,3$ and we have
used geometrised units such that $8\pi G=1=c$.

\section{Description of the Szekeres models}

\setcounter{equation}{0}

The average velocity of the matter in the Universe is identified with a unit
timelike vector field $u^{a}$ ($u^{a}u_{a}=-1$) tangent to the congruence of
worldlines of the fundamental (preferred) observers. The Einstein's Field
Equations (EFEs)\ for a pressureless perfect fluid can be expressed in terms
of the Ricci tensor as 
\begin{equation}
R_{ab}=\frac{\rho }{2}\left( u_{a}u_{b}+h_{ab}\right) +\Lambda g_{ab}
\label{EFE1}
\end{equation}
where $\rho $ is the energy density of the matter fluid as measured by the
comoving observers $u^{a}$, $\Lambda $ is the cosmological constant and $%
h_{ab}\equiv g_{ab}+u_{a}u_{b}$ is the projection tensor normally to $u^{a}$.

The matter velocity $u^{a}$ is geodesic (because of the vanishing of the
pressure) and we also assume that is irrotational (for an interesting
treatment of Szekeres' models with respect of tilted observers see \cite%
{Herrera:2012sc}). In the case of a \emph{vanishing cosmological constant},
the EFEs can be written in terms of the kinematical quantities of the
geodesic and irrotational timelike congruence by using the Ricci identities 
\begin{equation}
2u_{a;[bc]}=R_{dabc}u^{d}.  \label{RicciIdentity1}
\end{equation}%
Projecting equation (\ref{RicciIdentity1}) along the $u^{a}$ we obtain the
evolution equations \cite{Ellis:1998ct} 
\begin{equation}
\dot{H}=-H^{2}-\frac{1}{3}\sigma _{ab}\sigma ^{ab}-\frac{1}{6}\rho
\label{evolutionexpansion}
\end{equation}%
\begin{equation}
\dot{\sigma}_{ab}=-2H\sigma _{ab}-\sigma _{a}^{\hspace{0.2cm}c}\sigma
_{cb}-E_{ab}+\frac{1}{3}\left( \sigma _{cd}\sigma ^{cd}\right) h_{ab}
\label{evolutionshear}
\end{equation}%
for the overall volume expansion (or Hubble parameter) rate $3H=\theta
=u_{a;b}h^{ab}$ and the shear tensor $\sigma _{ab}=h_{a}^{\hspace{0.15cm}%
k}h_{b}^{\hspace{0.15cm}l}\left[ u_{\left( k;l\right) }-\frac{\theta }{3}%
h_{kl}\right] $ which describes the rate of distortion of the rest space of $%
u^{a}$ in different directions (i.e. the change of its shape). On the other
hand projecting normally to $u^{a}$, equation \ (\ref{RicciIdentity1}) leads
to the \emph{constraint} equations (i.e. the spatial divergence and curl
equations or equivalently the fully projected derivatives normal to the
timelike congruence $u^{a}$) 
\begin{equation}
h_{ac}\sigma _{\hspace{0.3cm};k}^{kc}=2h_{a}^{\hspace{0.2cm}k}H_{;k}
\label{divergenceshear}
\end{equation}%
\begin{equation}
h^{l(a}\epsilon ^{b)mns}\sigma _{nl;m}u_{s}=0  \label{shearconstraint}
\end{equation}%
where $\epsilon ^{abcd}$ is the 4-dimensional volume element and a dot
denotes covariant differentiation along the direction of the vector field $%
u^{a}$.

In order to obtain a closed set of equations we use the Bianchi identities 
\begin{equation}
R_{ab[cd;e]}=0.  \label{BianchiIdentities1}
\end{equation}%
The associated $u^{a}$ and $h_{ab}-$projections of (\ref{BianchiIdentities1}%
) give the energy and momentum conservation equations plus two evolution and
two constraints equations for the electric and magnetic part of the Weyl
tensor which for the case of the Szekeres models take the form (we recall
that the magnetic part of the Weyl tensor vanishes \cite{Barnes-Rowlingson})

\begin{equation}
\dot{E}_{ab}=-\frac{1}{2}\rho \sigma _{ab}-3HE_{ab}+3E_{(a}^{\hspace{0.2cm}%
c}\sigma _{b)c}-\left( E_{cd}\sigma ^{cd}\right) h_{ab}
\label{evolutionelectric}
\end{equation}%
\begin{equation}
\dot{\rho}=-3\rho H  \label{energyconservation}
\end{equation}%
\begin{equation}
h_{ac}E_{\hspace{0.3cm};k}^{kc}=\frac{1}{3}h_{a}^{\hspace{0.2cm}k}\rho _{;k}
\label{electricdivergence}
\end{equation}%
\begin{equation}
h^{l(a}\epsilon ^{b)mns}E_{nl;m}u_{s}=0  \label{electricconstraint}
\end{equation}%
\begin{equation}
\epsilon ^{abcd}E_{bk}\sigma _{c}^{\hspace{0.2cm}k}u_{d}=0.
\label{magneticconstraint}
\end{equation}%
In the above system of equations (\ref{evolutionexpansion})-(\ref%
{evolutionshear}) and (\ref{evolutionelectric})-(\ref{magneticconstraint})
we observe that the spatial divergence and curl equations (\ref%
{divergenceshear})-(\ref{shearconstraint}) and (\ref{electricdivergence})-(%
\ref{electricconstraint}) have been decoupled from the evolution equations
of the kinematical and dynamical variables. As a result, the evolution of
Szekeres models is completely independent from the spatial variations of the
physical variables and is described by a set of first-order ordinary
differential equations (ODEs).

It is convenient to employ a set of three mutually orthogonal and unit
spacelike vector fields $\left\{ x^{a},y^{a},z^{a}\right\} $ which can be
uniquely\emph{\ }chosen as \emph{eigenvectors} of the shear tensor. Equation
(\ref{magneticconstraint}) implies that the shear $\sigma _{ab}$ and the
electric part $E_{ab}$ tensors \emph{commute} hence they share the
eigenframe $\left\{ x^{a},y^{a},z^{a}\right\} $. We write

\begin{equation}
E_{ab}=E_{1}x_{a}x_{b}+E_{2}y_{a}y_{b}+E_{3}z_{a}z_{b}
\label{electricparteigen}
\end{equation}%
\begin{equation}
\sigma _{ab}=\sigma _{1}x_{a}x_{b}+\sigma _{2}y_{a}y_{b}+\sigma
_{3}z_{a}z_{b}  \label{sheareigen}
\end{equation}%
where $E_{\alpha },\sigma _{\alpha }$ are the associated eigenvalues
satisfying the trace-free conditions 
\begin{equation}
\sum_{\alpha }E_{\alpha }=\sum_{\alpha }\sigma _{\alpha }=0.
\label{trace-freecondition}
\end{equation}%
Furthermore it has been shown that each of $\left\{
x^{a},y^{a},z^{a}\right\} $ is hypersurface orthogonal \cite%
{Barnes-Rowlingson} i.e. 
\begin{equation}
x_{[a}x_{b;c]}=y_{[a}y_{b;c]}=z_{[a}z_{b;c]}=0.
\label{hypersurfaceorthogonal}
\end{equation}%
Due to the hypersurface property of the timelike and spacelike vector
fields, each pair $\left\{ u^{a},x^{a}\right\} $, $\left\{
u^{a},y^{a}\right\} $, $\left\{ u^{a},z^{a}\right\} $ generates a 2d
integrable submanifold of $\mathcal{M}$ represented by the 2-form e.g. $%
\mathbf{F}_{x}=\mathbf{u}\wedge \mathbf{x}$. It turns out that at each point
the spacetime manifold admits orthogonal 2-surfaces spanned by the vectors $%
\{y^{a},z^{a}\}$ with a corresponding surface element the dual 2-form $%
\mathbf{\tilde{F}}_{x}$ (or equivalently the simple bivector $\mathbf{C}_{x}=%
\mathbf{y}\wedge \mathbf{z}$).

It should be emphasized that the above considerations hold, in general, for
homogeneous and inhomogeneous Petrov type I silent models. However, in the
present work we are interested in spatially inhomogeneous setups in which
case only the Petrov type D subclass of models exists \cite%
{Apostolopoulos:2006zn} and is described by the Szekeres family of solutions
satisfying 
\begin{equation}
E_{2}=E_{3}\Leftrightarrow \sigma _{2}=\sigma _{3}.
\label{SzekeresConditions}
\end{equation}%
In this case, local orthogonal coordinates can be found such that the metric
can be written \cite{Szekeres:1974ct} 
\begin{equation}
ds^{2}=g_{ab}dx^{a}dx^{b}=-dt^{2}+e^{2A}dr^{2}+e^{2B}\left(
dy^{2}+dz^{2}\right) .  \label{SIISmetric}
\end{equation}%
The \emph{general} solution of the resulting set of EFEs for a dust fluid,
implies that the smooth functions $A(t,r,y,z)$, $B(t,r,y,z)$ have the form 
\begin{equation}
e^{A}=\frac{S_{,r}-S\left( \ln E\right) _{,r}}{\sqrt{f+\epsilon }},\hspace{%
0.4cm}e^{B}=\frac{S}{E}.  \label{Cartesian1}
\end{equation}%
We shall see in section 4 that the value of $\epsilon =1,-1,0$ describes the
topology of the 2d hypersurface $\mathcal{X}$ (i.e. the distribution $%
\mathbf{F}_{x}=\mathbf{0}$), the function $S(t,r)$ is \emph{isotropic}
representing a \emph{generalized scale factor} that corresponds locally to
the \emph{length scale of the dust cloud} satisfying the equation of motion 
\begin{equation}
\left( \dot{S}\right) ^{2}=\frac{2\mathcal{M}_{1}}{S}+f
\label{GeneralizedFriedm0}
\end{equation}%
where $\mathcal{M}_{1}(r)$, $f(r)$ are functions of the radial coordinate
and $E(r,y,z)$ controls the 2d anisotropy of $\mathcal{X}$ 
\begin{equation}
E=A(r)\left( y^{2}+z^{2}\right) +B(r)y+C(r)z+D(r).
\label{FunctionECartesian}
\end{equation}%
The functions $A(r),B(r),C(r)$ and $D(r)$ are subjected to the algebraic
constraint 
\begin{equation}
4AD-B^{2}-C^{2}=\epsilon .  \label{Constraint}
\end{equation}%
The relation (\ref{Constraint}) implies that the \textquotedblleft
anisotropy\textquotedblright\ function $E(r,y,z)$ can be written \cite%
{Hellaby:1996zz, Hellaby:2002nx, Apostolopoulos:2016nno, Georg:2017tpz} 
\begin{equation}
E(r,y,z)=\frac{V}{2}\left\{ \left[ \frac{y-Y(r)}{V}\right] ^{2}+\left[ \frac{%
z-Z(r)}{V}\right] ^{2}+\epsilon \right\}  \label{FunctionE1}
\end{equation}%
where $Y(r)=-B\cdot V$, $Z(r)=-C\cdot V$ and $V(r)$ are functions of $r$.
With these identifications the Szekeres spacetime takes the form 
\begin{equation}
ds^{2}=-dt^{2}+S^{2}\left\{ \frac{\left[ \left( \ln S/E\right) ^{\prime }%
\right] ^{2}}{\epsilon +f}dr^{2}+\frac{4\left( dy^{2}+dz^{2}\right) }{V^{2}%
\left[ \left( \frac{y-Y}{V}\right) ^{2}+\left( \frac{z-Z}{V}\right)
^{2}+\epsilon \right] ^{2}}\right\}
\label{SzekeresMetricHellabyKrasinskiForm1}
\end{equation}%
The local form of the metric (\ref{SzekeresMetricHellabyKrasinskiForm1})
manifests the constancy of the curvature of the distribution $\mathcal{X}$
which in turn restricts the coordinate dependence of several crucial
kinematical/dynamical quantities. Furthermore eq. (\ref%
{SzekeresMetricHellabyKrasinskiForm1}) has the additional advantage to
exhibit in a transparent and natural way, important subfamilies with higher
degree of symmetries. For example, the spherically/hyperbolic/plane (LRS)\
symmetric model follows from the choice $B=C=0$ and $A=\epsilon D=1/2$ ($%
\epsilon \neq 0$), $D=0$ ($\epsilon =0$).

The Szekeres models fall into two classes (I and II) depending on the radial
dependence of the metric functions. Apart from (\ref%
{SzekeresMetricHellabyKrasinskiForm1}), several other forms of the metric
have been used in the literature describing e.g. quasi-spherical collapsing
clouds of dust, generalizations of the Kantowski-Sachs LRS\ cosmological
models or to emphasize similarities of the dynamics of the Szekeres models
with the linear perturbations of the Friedmann-Robertson-Walker model \cite%
{Bonnor1, BoonorTomimura, Goode-Wainwright}. Because we are interested to
use Szekeres models in order to analyze the effect of local inhomogeneities
in the cosmological observations it is also convenient to choose
(pseudo-)spherical coordinates to express the local form of the metric \cite%
{Nolan-Debnath}. In this case we get 
\begin{equation}
ds^{2}=-dt^{2}+\frac{E^{2}\left[ \left( SE^{-1}\right) _{,r}\right] ^{2}}{%
f+\epsilon }dr^{2}+\frac{S^{2}}{E^{2}}\left( d\theta ^{2}+\Sigma ^{2}d\phi
^{2}\right) .  \label{Spherical1}
\end{equation}%
The function $E(r,\theta ,\phi )$ reads 
\[
E=(A-\epsilon D)\Lambda (\epsilon ,\theta )+B\Sigma (\epsilon ,\theta )\cos
\phi +C\Sigma (\epsilon ,\theta )\sin \phi +\epsilon ^{2}(A+\epsilon D) 
\]%
where $\Sigma (\epsilon ,\theta )=\left( \sin \theta ,\sinh \theta ,\theta
\right) $, $\Lambda (\epsilon ,\theta )=\left( \cos \theta ,\cosh \theta
,\theta ^{2}\right) $ for $\epsilon =1,-1,0$ respectively.

\section{1+1+2 decomposition}

\setcounter{equation}{0}

The constraint equations (\ref{divergenceshear}), (\ref{shearconstraint}), (%
\ref{electricdivergence}) and (\ref{electricconstraint}) can be conveniently
reformulated \emph{in terms of the corresponding kinematical quantities of
the spacelike congruences} generated from each of the spacelike vector
fields $\left\{ x^{a},y^{a},z^{a}\right\} $. In particular, it has been
shown that the first derivatives of the spacelike eigenvector fields are
decomposed as \cite{Apostolopoulos:2006zn} 
\begin{equation}
x_{a;b}=\mathcal{T}_{ab}(\mathbf{x})+\frac{\mathcal{E}_{x}}{2}p_{ab}(\mathbf{%
x})+\left( x_{a}\right) ^{\ast }x_{b}  \label{spacelikex}
\end{equation}%
\begin{equation}
y_{a;b}=\mathcal{T}_{ab}(\mathbf{y})+\frac{\mathcal{E}_{y}}{2}p_{ab}(\mathbf{%
y})+y_{a}^{\prime }y_{b}  \label{spacelikey}
\end{equation}%
\begin{equation}
z_{a;b}=\mathcal{T}_{ab}(\mathbf{z})+\frac{\mathcal{E}_{z}}{2}p_{ab}(\mathbf{%
z})+\tilde{z}_{a}z_{b}  \label{spacelikez}
\end{equation}%
where \footnote{%
In \cite{Clarkson:2002jz} the 1+1+2 covariant formalism has been exploited
in studying covariant perturbations of a Schwarzschild black hole using a
completely different notation of the various quantities.} \cite%
{Spacelike-Congruences-Set-Of-Papers} 
\begin{equation}
\mathcal{E}_{x}=x_{a;b}p^{ab}(\mathbf{x}),\hspace{0.3cm}\mathcal{E}%
_{y}=y_{a;b}p^{ab}(\mathbf{y}),\hspace{0.3cm}\mathcal{E}_{z}=z_{a;b}p^{ab}(%
\mathbf{z})  \label{spatialexpansions}
\end{equation}%
\begin{eqnarray}
\mathcal{T}_{ab}(\mathbf{x}) &=&\alpha \left( y_{a}y_{b}-z_{a}z_{b}\right) ,%
\hspace{0.3cm}\mathcal{T}_{ab}(\mathbf{y})=\beta \left(
z_{a}z_{b}-x_{a}x_{b}\right)  \nonumber \\
&&%
\begin{tabular}{l}
$\mathcal{T}_{ab}(\mathbf{z})=\gamma \left( x_{a}x_{b}-y_{a}y_{b}\right) $%
\end{tabular}
\label{spatialshears}
\end{eqnarray}%
are the rates of the 2d \emph{surface area spatial expansion} and of the
(traceless) shear tensor of the spacelike congruences respectively and we
have used the notation 
\begin{eqnarray}
\left( K_{a...}\right) ^{\ast } &\equiv &K_{a...;k}x^{k},\hspace{0.5cm}%
K_{a...}^{\prime }\equiv K_{a...;k}y^{k},  \nonumber \\
&&%
\begin{tabular}{l}
$\left( K_{a...}\right) ^{\symbol{126}}\equiv K_{a...;k}z^{k}.$%
\end{tabular}
\label{spatialderivatives}
\end{eqnarray}%
Using the Ricci identity (\ref{RicciIdentity1}) it is straightforward to
show the following \emph{commutation relations} for every scalar quantity $S$
\begin{equation}
\left( {S}^{\ast }\right) ^{\cdot }=\left( \dot{S}\right) ^{\ast }-%
{S}^{\ast }\left( \sigma _{1}+H\right)  \label{xcommutaionrelation}
\end{equation}%
\begin{equation}
\left( S^{\prime }\right) ^{\cdot }=\left( \dot{S}\right) ^{\prime
}-S^{\prime }\left( \sigma _{2}+H\right)  \label{ycommutaionrelation}
\end{equation}%
\begin{equation}
\left( \tilde{S}\right) ^{\cdot }=\left( \dot{S}\right) ^{\symbol{126}}-%
\tilde{S}\left( \sigma _{3}+H\right) .  \label{zcommutaionrelation}
\end{equation}%
The projection tensors $p_{ab}(\mathbf{e})$ associated with $\mathbf{e}=\{%
\mathbf{x},\mathbf{y},\mathbf{z}\}$ are given by 
\begin{equation}
p_{ab}(\mathbf{e})\equiv g_{ab}+u_{a}u_{b}-e_{a}e_{b}=h_{ab}-e_{a}e_{b}
\label{ProjectorTensor1}
\end{equation}%
and are identified with the corresponding metrics of the 2d spaces $\mathcal{%
X}$ (\emph{screen spaces}) orthogonal to each vector field of the pairs $%
\left\{ u^{a},x^{a}\right\} $, $\left\{ u^{a},y^{a}\right\} $, $\left\{
u^{a},z^{a}\right\} $ at any spacetime event. Each of the screen spaces is
an assembly of 2-surfaces which are different with each other. However, due
to the vanishing of the spatial twist $\mathcal{R}_{ab}(\mathbf{e})=p_{a}^{%
\hspace{0.15cm}k}p_{b}^{\hspace{0.15cm}l}e_{[k;l]}$ and Greenberg vector $%
N_{a}=p_{a}^{\hspace{0.15cm}k}\mathcal{L}_{\mathbf{u}}e_{k}$ \emph{the
screen space }$\mathcal{X}$\emph{\ is a genuine 2d surface which is a
submanifold of the observers' instantaneous rest space} \cite%
{Apostolopoulos:2006zn}.

We shall restrict our considerations to the Szekeres models (i.e. Spatially
Inhomogeneous Irrotational Silent (SIIS) models of Petrov type D) for which $%
\mathcal{T}_{ab}(\mathbf{x})=0=\alpha $ \cite{Apostolopoulos:2006zn}. The
evolution equations (\ref{evolutionshear}), (\ref{evolutionelectric}), (\ref%
{energyconservation}) and the divergence equations (\ref{divergenceshear}), (%
\ref{electricdivergence}) are written in the form

\begin{equation}
\dot{H}=-H^{2}-2\sigma _{3}^{2}-\frac{1}{6}\rho  \label{EvolutionHubble}
\end{equation}

\begin{equation}
\dot{\sigma}_{3}=-2H\sigma _{3}+\left( \sigma _{3}\right) ^{2}-E_{3}
\label{sheareigenvaluesevolution}
\end{equation}%
\begin{equation}
\dot{E}_{3}=-\frac{1}{2}\rho \sigma _{3}-3\left( H+\sigma _{3}\right) E_{3}
\label{electriceigenvaluesevolution}
\end{equation}%
\begin{equation}
\dot{\rho}=-\rho \theta  \label{energyconservation2}
\end{equation}%
\begin{eqnarray}
\left( \rho \right) ^{\ast } &=&-3\left[ 2\left( E_{3}\right) ^{\ast }+3E_{3}%
\mathcal{E}_{x}\right]  \nonumber \\
\left( H\right) ^{\ast } &=&-\frac{1}{2}\left[ 2\left( \sigma _{3}\right)
^{\ast }+3\sigma _{3}\mathcal{E}_{x}\right]  \label{xdivergence}
\end{eqnarray}%
\begin{equation}
\rho ^{\prime }=6E_{3}\left( \frac{\mathcal{E}_{y}}{2}-\beta \right) ,%
\hspace{0.5cm}\tilde{\rho}=6E_{3}\left( \frac{\mathcal{E}_{z}}{2}+\gamma
\right)  \label{yzdivergencedensity}
\end{equation}%
\begin{equation}
H^{\prime }=\left( \frac{\mathcal{E}_{y}}{2}-\beta \right) \sigma _{3},%
\hspace{0.5cm}\tilde{H}=\left( \frac{\mathcal{E}_{z}}{2}+\gamma \right)
\sigma _{3}.  \label{yzdivergenceexpansion}
\end{equation}%
Projecting (\ref{shearconstraint}) and (\ref{electricconstraint}) along $%
y^{a}z^{b}$, $z^{a}x^{b}$ and $x^{a}y^{b}$, the shear and electric part
constraints can be expressed as spatial variations of the corresponding
eigenvalues along the individual spacelike curve 
\begin{equation}
\sigma _{3}^{\prime }=-\left( \frac{\mathcal{E}_{y}}{2}-\beta \right) \sigma
_{3},\hspace{0.5cm}\left( \sigma _{3}\right) ^{\symbol{126}}=-\left( \frac{%
\mathcal{E}_{z}}{2}+\gamma \right) \sigma _{3}  \label{yzshearconstraint}
\end{equation}%
\begin{equation}
E_{3}^{\prime }=-\left( \frac{\mathcal{E}_{y}}{2}-\beta \right) E_{3},%
\hspace{0.5cm}\left( E_{3}\right) ^{\symbol{126}}=-\left( \frac{\mathcal{E}%
_{z}}{2}+\gamma \right) E_{3}  \label{yzelectricconstraint}
\end{equation}%
The above set of equations must be augmented with the Ricci identity and the
Jacobi identities for the orthonormal triad $\left\{
x^{a},y^{a},z^{a}\right\} $. In particular, the $u^{a}-$projected Ricci
identity gives evolution equations of the kinematical quantities of the
spacelike congruences \cite{Apostolopoulos:2006zn} 
\begin{equation}
\left( \mathcal{E}_{x}\right) ^{\cdot }=-\mathcal{E}_{x}\left( \sigma
_{3}+H\right)  \label{evolutionxcongruence}
\end{equation}%
\begin{equation}
\left( \frac{\mathcal{E}_{y}}{2}+\beta \right) ^{\cdot }=-\left( \frac{%
\mathcal{E}_{y}}{2}+\beta \right) \left( \sigma _{3}+H\right)
\label{evolutionycongruence1}
\end{equation}%
\begin{equation}
\left( \frac{\mathcal{E}_{y}}{2}-\beta \right) ^{\cdot }=-\left( \frac{%
\mathcal{E}_{y}}{2}-\beta \right) \left( \sigma _{1}+H\right)
\label{evolutionycongruence2}
\end{equation}%
\begin{equation}
\left( \frac{\mathcal{E}_{z}}{2}+\gamma \right) ^{\cdot }=-\left( \frac{%
\mathcal{E}_{z}}{2}+\gamma \right) \left( \sigma _{1}+H\right)
\label{evolutionzcongruence1}
\end{equation}%
\begin{equation}
\left( \frac{\mathcal{E}_{z}}{2}-\gamma \right) ^{\cdot }=-\left( \frac{%
\mathcal{E}_{z}}{2}-\gamma \right) \left( \sigma _{3}+H\right) .
\label{evolutionzcongruence2}
\end{equation}%
The constraint equations arise from the spatial projections of the trace
part of the Ricci identity of each spacelike vector field. We note that an
appropriate linear combination of the constraint equation leads to the
corresponding Jacobi identities or, equivalently, to the twist-free property 
$\mathcal{R}=0$ of the spacelike congruences 
\begin{equation}
\left( \mathcal{E}_{x}\right) ^{\prime }=0,\hspace{0.5cm}\left( \mathcal{E}%
_{x}\right) ^{\symbol{126}}=0  \label{spatialXRicciIdentity}
\end{equation}%
\begin{equation}
\left( \frac{\mathcal{E}_{y}}{2}+\beta \right) ^{\ast }=-\beta \mathcal{E}%
_{x},\hspace{0.3cm}\left( \frac{\mathcal{E}_{y}}{2}-\beta \right) ^{\symbol{%
126}}=2\beta \left( \frac{\mathcal{E}_{z}}{2}+\gamma \right)
\label{spatialYRicciIdentity}
\end{equation}%
\begin{equation}
\left( \frac{\mathcal{E}_{z}}{2}-\gamma \right) ^{\ast }=\gamma \mathcal{E}%
_{x},\hspace{0.3cm}\left( \frac{\mathcal{E}_{z}}{2}+\gamma \right) ^{\prime
}=-2\gamma \left( \frac{\mathcal{E}_{y}}{2}-\beta \right) .
\label{spatialZRicciIdentity}
\end{equation}%
Furthermore, from the $h_{a}^{\hspace{0.2cm}k}-$projected Ricci identity for
the vector fields $\left\{ x^{a},y^{a},z^{a}\right\} $ we obtain spatial
propagation equations of the spacelike expansion and shear rates \cite%
{Apostolopoulos:2006zn}. This set of equations completely characterizes the
dynamics of the Szekeres models in terms of the kinematical quantities of
the timelike and spacelike congruences. Nevertheless, as we shall see in the
next section, a more detailed analysis of the above set of equations reveals
a number of \textquotedblleft hidden\textquotedblright\ properties of the
Szekeres models with sound geometrical and physical usefulness.

\section{Geometrical and dynamical covariant analysis}

\setcounter{equation}{0}

The fact that the dust fluid velocity is geodesic and irrotational, implies
that the hypersurfaces $t=$const. form an integrable submanifold $\mathcal{S}
$ of $\mathcal{M}$ with metric $h_{ab}$ and extrinsic curvature $\Theta
_{ab}=u_{(a;b)}$. Then Gauss equation implies

\begin{equation}
^{3}R_{abcd}=h_{a}^{\hspace{0.2cm}i}h_{b}^{\hspace{0.2cm}j}h_{c}^{\hspace{%
0.2cm}k}h_{d}^{\hspace{0.2cm}l}R_{ijkl}+2\Theta _{a[d}\Theta _{c]b}
\label{GaussEquation1}
\end{equation}%
i.e. the 3d curvature tensor $^{3}R_{abcd}$ of $\mathcal{S}$ is expressed in
terms of the projected curvature tensor of the spacetime plus extrinsic
curvature corrections. The 3d curvature tensor is completely determined by
the Ricci tensor and the scalar curvature of the 3-spaces $\mathcal{S}$
which, by virtue of (\ref{GaussEquation1}), are given by 
\begin{equation}
^{3}R_{ab}=E_{ab}-H\sigma _{ab}+\sigma _{a}^{\hspace{0.2cm}k}\sigma _{kb}+%
\frac{2}{3}\left( \rho -3H^{2}\right) h_{ab}  \label{3RicciTensor1}
\end{equation}%
\begin{equation}
^{3}R=2\left[ \rho +3\left( \sigma _{3}^{2}-H^{2}\right) \right] .
\label{FriedmannEquation}
\end{equation}%
Essentially the last relation represents the generalized Friedmann equation
in the Szekeres models and as we shall see below, it coincides with the
equation of motion of the sheets constituting the clouds of dust.

It is interesting to remark that \emph{when} the following condition holds 
\begin{equation}
\left( H+\sigma _{3}\right) ^{2}=2\left( \frac{\rho }{6}+E_{3}\right) .
\label{FlatKVFCondition1}
\end{equation}%
then, using eqs. (\ref{3RicciTensor1})-(\ref{FriedmannEquation}), the 3d
Ricci tensor can be written 
\begin{equation}
^{3}R_{ab}=\frac{^{3}R}{4}\left( x_{a}x_{b}+h_{ab}\right) .
\label{3RicciTensor2}
\end{equation}%
Taking the spatial divergence of equation (\ref{3RicciTensor2}) and using
the $h_{a}^{\hspace{0.2cm}i}h^{jk}\hspace{0.1cm}^{3}R_{ij;k}=\frac{1}{2}%
h_{a}^{\hspace{0.2cm}i}\hspace{0.1cm}^{3}R_{;i}$ we obtain 
\begin{equation}
^{3}R\mathcal{E}_{x}=0.  \label{3ScalarExpansionx1}
\end{equation}%
Equation (\ref{3ScalarExpansionx1}) means that when equation (\ref%
{FlatKVFCondition1}) holds then either \emph{the spatial slices are flat}
i.e. $^{3}R_{ab}=0$ ($\Rightarrow E_{3}=(\sigma _{3}+H)\sigma _{3}$) or
there exists a \emph{spacelike KVF parallel} to the eigenvector $x^{a}$. It
can be seen that for the models $\mathcal{E}_{x}\neq 0$ the condition (\ref%
{FlatKVFCondition1}) is necessary and sufficient for the flatness of the 3d
space $\mathcal{S}$.

Because Szekeres models represent a collapsing/expanding dust matter
configuration, it will be helpful to investigate the influence of the matter
content into the curvature of the 2d screen space $\mathcal{X}(\mathbf{x})$.
This can be achieved following a similar approach as before. The
hypersurface $t=$const. and $r=$const. is an integrable submanifold of $%
\mathcal{S}$ since the associated spacelike vorticity tensor and the
Greenberg vector vanish \cite{Apostolopoulos:2006zn}. In particular, the
corresponding Gauss equation for the submanifolds $\mathcal{S}$ reads

\begin{equation}
^{2}R_{abcd}=p_{a}^{\hspace{0.2cm}i}p_{b}^{\hspace{0.2cm}j}p_{c}^{\hspace{%
0.2cm}k}p_{d}^{\hspace{0.2cm}l}\hspace{0.2cm}^{3}R_{ijkl}+2K_{a[c}K_{d]b}
\label{GaussEquation2}
\end{equation}%
where $K_{ab}=p_{a}^{\hspace{0.2cm}i}p_{b}^{\hspace{0.2cm}j}x_{(i;j)}\equiv 
\bar{\nabla}_{(b}x_{a)}$ is the extrinsic curvature, \textquotedblleft $\bar{%
\nabla}_{a}$\textquotedblright\ denotes the proper covariant derivative and $%
^{2}R_{abcd}$ is the curvature tensor of the screen space $\mathcal{X}(%
\mathbf{x})$ with metric $p_{ab}(\mathbf{x})=h_{ab}-x_{a}x_{b}$. Contracting
twice equation (\ref{GaussEquation2}) and using (\ref{FriedmannEquation}) we
obtain 
\begin{equation}
^{2}R=4\left( \frac{\rho }{6}+E_{3}\right) -2\left( H+\sigma _{3}\right)
^{2}+\frac{\mathcal{E}_{x}^{2}}{2}.  \label{2ScalarCurvature1}
\end{equation}%
The last equation can be regarded as the \emph{equation of motion of the
dust shells}. In particular eq. (\ref{2ScalarCurvature1}) shows how the
scalar curvature of the 2-space $\mathcal{X}(\mathbf{x})$ is affected by the
kinematics/dynamics and vice versa. For example when the condition (\ref%
{FlatKVFCondition1}) holds, the curvature of the 2d space $\mathcal{X}(%
\mathbf{x})$ is completely determined by the spacelike expansion $\mathcal{E}%
_{x}$. It follows that the flat 3d space $\mathcal{S}$ (for the models with $%
\mathcal{E}_{x}\neq 0$) is foliated by 2d sheets of positive curvature.
Furthermore for models admitting a spacelike KVF parallel to the eigenvector 
$x^{a}$ ($\mathcal{E}_{x}=0$), eq. (\ref{FlatKVFCondition1}) is the
necessary and sufficient condition for the flatness ($^{2}R=0$) of the 2d
screen space $\mathcal{X}(\mathbf{x})$.

Equation (\ref{2ScalarCurvature1}) suggests that one must analyze the role
of the quantities 
\begin{equation}
w_{1}\equiv H+\sigma _{3},\hspace{0.5cm}w_{2}\equiv \frac{\rho }{6}+E_{3}
\label{w1w2quantities}
\end{equation}%
in manufacturing the Szekeres models.

Let us consider first the quantity $w_{1}$ and observe that can be written
as 
\begin{equation}
w_{1}=H+\sigma _{3}=\frac{1}{2}u_{a;b}p^{ab}  \label{w1analysis1}
\end{equation}%
i.e. $w_{1}$ \emph{is the trace of the first derivatives of the
four-velocity projected in the screen space} $\mathcal{X}(\mathbf{x})$. This
implies that, in complete analogy with the meaning of $H$, the quantity $%
w_{1}$ represents the \emph{rate of change or the expansion rate of the
surface area} of the 2-space $\mathcal{X}(\mathbf{x})$. We shall refer to it
as \emph{area expansion}. We point out that, although Szekeres models are,
in general, quasi-symmetric (i.e. no isotropic 2d sections exist), the
surface area expands \emph{isotropically} (due to equations (\ref%
{yzdivergenceexpansion})-(\ref{yzshearconstraint})) 
\begin{equation}
\left( w_{1}\right) ^{\prime }=\left( w_{1}\right) ^{\symbol{126}}=0.
\label{VanishingyzDerivatives1}
\end{equation}%
It would be convenient to define an \emph{average length scale} $\ell $
according to 
\begin{equation}
w_{1}\equiv \frac{\dot{\ell}}{\ell }.  \label{LengthScale1}
\end{equation}%
We note that the intuitive definition (\ref{LengthScale1}) is dictated
directly from (\ref{xdivergence}).

Taking into account the above definitions, the length $\ell $ completely
determines the surface area of $\mathcal{X}(\mathbf{x})$ which scales $\sim
\ell ^{2}$ as the 2-spaces $\mathcal{X}(\mathbf{x})$ evolve. The conditions
of isotropy (\ref{VanishingyzDerivatives1}) allow us to define covariantly
the scalar $S$ according

\begin{equation}
\frac{\dot{\ell}}{\ell }=\frac{\dot{S}}{S}  \label{GeneralScaleFactor1}
\end{equation}%
where $S^{\prime }=\tilde{S}=0$ such that the average length can be written
as $\ell =S\cdot N$ with $\dot{N}=0$. The scalar $S$ can be seen as
auxiliary quantity without obvious geometrical and physical meaning.
Nevertheless we will show covariantly and computationally in subsequent
paragraphs that $S$ satisfies the same dynamical equation (\ref%
{GeneralizedFriedm0}) as the standard areal \textquotedblleft
radius\textquotedblright\ of the Szekeres family. Therefore, we can argue
that the physical interpretation of $S$ is that it represents a \emph{%
generalized scale factor} and corresponds, for each $t=$const. and $r=$%
const., to the \emph{length scale of the dust cloud }(or the \emph{radius
shell }in the case $^{2}R(t=$const.$,r=$const.$)>0$). Furthermore the
covariant (static) scalar $N$ exhibits the anisotropy of the 2d screen space
and the quasi-symmetrical feature of the Szekeres models (formally, in the
coordinates adapted in (\ref{SzekeresMetricHellabyKrasinskiForm1}), it
corresponds to $E^{-1}$).

Similarly, the spatial ($x-$)change of $\ell $ is controlled by the
expansion rate of the spacelike congruence generated by the vector filed $%
x^{a}$ 
\begin{equation}
\mathcal{E}_{x}=2\frac{\left( \ell \right) ^{\ast }}{\ell }.
\label{LengthScale2}
\end{equation}%
Consequently, for the Szekeres models that do not admit a spacelike KVF
parallel to $x^{a}$, the condition $\mathcal{E}_{x}=0\Leftrightarrow \left(
\ell \right) ^{\ast }=0$ implies that the curves of the spacelike congruence
converge to a single curve. Physically this means that different dust
regions collide and stick together in a single sheet allowing the \emph{%
formation of a crossing singularity} \cite{Hellaby:2002nx}. This situation
is similar to the generation of a shell-crossing singularity in the case of
the Locally Rotationally Symmetric (LRS) Spatially Inhomogeneous
Irrotational Silent models (LT solution corresponds to the positive 2d
curvature subclass).

On the other hand, it can be easily seen that the quantity $w_{2}$ is
defined as the sum of the total energy density $\rho $ of the dust fluid
plus the contribution from the free gravitational field (encoded in the
eigenvalue $E_{3}$ of the electric part of the Weyl tensor) and shares the
same \textquotedblleft isotropic\textquotedblright\ property with $w_{1}$ 
\begin{equation}
\left( w_{2}\right) ^{\prime }=\left( w_{2}\right) ^{\symbol{126}}=0.
\label{VanishingyzDerivatives2}
\end{equation}%
Using the first of (\ref{xdivergence}) and the relation (\ref{LengthScale2})
we get 
\begin{equation}
\left( \frac{\rho }{6}\right) ^{\ast }\ell ^{3}+\left( E_{3}\ell ^{3}\right)
^{\ast }=0\Rightarrow \left( w_{2}\right) ^{\ast }=\left[ \left( \frac{\rho 
}{6}+E_{3}\right) \ell ^{3}\right] ^{\ast }=\frac{\rho \ell ^{2}\left( \ell
\right) ^{\ast }}{2}.  \label{Auxiliary1}
\end{equation}%
Therefore we can rewrite $w_{2}$ in the following useful form 
\begin{equation}
w_{2}\equiv \frac{\rho }{6}+E_{3}=\frac{\mathcal{M}_{\mathrm{eff}}}{\ell ^{3}%
}  \label{ComovingMass1}
\end{equation}%
where 
\begin{equation}
\left( \mathcal{M}_{{\mathrm{eff}}}\right) ^{\ast }=\frac{\rho \ell
^{2}\left( \ell \right) ^{\ast }}{2}  \label{ComovingMass2}
\end{equation}%
is interpreted as the \emph{effective gravitational mass} (EGM) contained in
the dust cloud with length scale $\ell $.

This interpretation is strictly correct only in the quasi-spherical case $%
^{2}R(t=$const.$, r=$const.$)>0$ however, for the sake of simplicity, we
shall continue to refer it as EGM. It should be emphasized that equations (%
\ref{VanishingyzDerivatives1}), (\ref{VanishingyzDerivatives2}) and (\ref%
{2ScalarCurvature1}) show that $^{2}R$ is constant along the $y^{a},z^{a}-$%
directions therefore the screen space is \emph{of constant curvature} and we
refer Szekeres models as quasi-symmetrical.

We note that in a series of interesting papers \cite{Sussman:2011bp,
Sussman:2012xc, Sussman:2013qya, Sussman:2013yq, Sussman:2015fwa} the LT and
Szekeres models are studied in terms of a similar set of quantities namely
the \textquotedblleft $q-$scalars\textquotedblright\ which are coordinate
independent functionals or \textquotedblleft quasi-local\textquotedblright\
variables (in the spirit of the integral definition of the quasi local
Misner-Sharp mass-energy function). However, our gauge-invariant approach
\textquotedblleft assigns\textquotedblright\ a \emph{unique} geometric or
dynamical identity to each covariant quantity and clarifies how the geometry
of the 2d screen space $\mathcal{X}(\mathbf{x})$ is affected from the
dynamics (and vice-versa). In particular, the conservation of the EGM holds
during the evolution \emph{irrespective} the curvature of the 2-space $%
\mathcal{X}(\mathbf{x})$. This can be shown by noticing that equations (\ref%
{electriceigenvaluesevolution}) and (\ref{energyconservation2}) imply $\dot{w%
}_{2}=-3w_{1}w_{2}$ therefore propagation of (\ref{ComovingMass1}) and use
of (\ref{LengthScale1}) gives 
\begin{equation}
\left( \mathcal{M}_{{\mathrm{eff}}}\right) ^{\cdot }=0.
\label{MassConservation1}
\end{equation}%
With these identifications, equation (\ref{2ScalarCurvature1}) is written 
\begin{equation}
\left( \dot{\ell}\right) ^{2}=\frac{2\mathcal{M}_{{\mathrm{eff}}}}{\ell }+%
\frac{\mathcal{E}_{x}^{2}-\hspace{0.2cm}2\hspace{0.2cm}^{2}R}{4}\ell ^{2}
\label{EquationOfMotion1}
\end{equation}%
and represents the \emph{equation of motion} (the \emph{Hamiltonian}) of the
sheets of the dust matter configuration.

In terms of the generalized scale factor $S$ and the arbitrary scalar $%
\mathcal{M}_{1}=\mathcal{M}_{{\mathrm{eff}}}N^{-3}$ ($\left( \mathcal{M}%
_{1}\right) ^{\bullet }=\mathcal{M}_{1}^{\prime }=\left( \mathcal{M}%
_{1}\right) ^{\mathbf{\symbol{126}}}=0$) the last equation is expressed as%
\begin{equation}
\left( \dot{S}\right) ^{2}=\frac{2\mathcal{M}_{1}}{S}+\frac{\mathcal{E}%
_{x}^{2}-\hspace{0.2cm}2\hspace{0.2cm}^{2}R}{4}S^{2}.
\label{EquationOfMotion2}
\end{equation}%
We observe that the evolution of the average length scale of the dust cloud%
\emph{\ }$S$ depends on the function $\mathcal{M}_{1}$, the \emph{effective
curvature term} and an arbitrary function $S_{0}$ ($\dot{S}%
_{0}=S_{0}^{\prime }=\tilde{S}_{0}=0$) which corresponds to the local time
of the big bang ($S=0$). On the other hand any density profile within the
Szekeres models is attributed to the arbitrary function $\mathcal{M}_{1}$
and the contribution of the free gravitational field $E_{3}$ satisfying
equation (\ref{yzelectricconstraint}).

At this point some comments regarding the coordinate representation of the
above considerations are in order. We are interested only in the case of the
quasi-spherical case i.e. $^{2}R>0$. The local form of the eigenvectors of
the shear and electric part tensors in the spherical coordinate chart $%
\left\{ t,r,\theta ,\phi \right\} $ are $x^{a}=b^{-1}\delta _{r}^{a}$, $%
y^{a}=S^{-1}E\delta _{\theta }^{a}$, $z^{a}=S^{-1}E\left( \sin \theta
\right) ^{-1}\delta _{\phi }^{a}$ and we have set $b=E\left( S/E\right)
_{,r}\left( f+\epsilon \right) ^{-1/2}$. A trivial computation gives 
\begin{equation}
\sigma _{1}+H=-2\sigma _{3}+H=\frac{b_{,t}}{b},\hspace{0.5cm}\sigma _{3}+H=%
\frac{S_{,t}}{S}  \label{FluidKinematical}
\end{equation}%
\begin{eqnarray}
\mathcal{E}_{x}=\frac{2\sqrt{f+\epsilon }}{S},\hspace{0.5cm}\frac{\mathcal{E}%
_{y}}{2} &+&\beta =\frac{E}{S}\left( \ln \frac{\sin \theta }{E}\right)
_{,\theta }=\frac{\ell ^{\prime }}{\ell },  \nonumber \\
&&  \nonumber \\
\frac{\mathcal{E}_{y}}{2}-\beta &=&\frac{E}{S}\frac{b_{,\theta }}{b}=\frac{%
b^{\prime }}{b}  \label{SpatialQuant1}
\end{eqnarray}%
\begin{eqnarray}
\frac{\mathcal{E}_{z}}{2}+\gamma &=&\frac{E}{S\sin \theta }\frac{b_{,\phi }}{%
b}=\frac{\tilde{b}}{b},  \nonumber \\
\frac{\mathcal{E}_{z}}{2}-\gamma &=&-\frac{E_{,\phi }}{S\sin \theta }=\frac{%
\tilde{\ell}}{\ell },  \label{SpatialQuant2}
\end{eqnarray}%
with $\ell =S\cdot E^{-1}\sin \theta $ (note that $N=E^{-1}\sin \theta $).
Interestingly, the linear combination of the $y,z-$kinematical variables has
also a similar geometrical interpretation, representing the spatial
variation of the average length scale $\ell $ of the screen space and the
scale $\hat{\ell}=b$ along the $y^{a},z^{b}$ curves.

The energy density of the dust fluid follows from equation (\ref%
{ComovingMass2}) (note that $\mathcal{M}_{1}(r)$ is now a function of the
radial coordinate) 
\begin{equation}
\rho =\frac{2}{S^{3}}\frac{\left( \mathcal{M}_{1}\right) _{,r}+3\mathcal{M}%
_{1}E_{,r}}{\left( \ln E^{-1}S\right) _{,r}}.  \label{EnergyDensity2}
\end{equation}%
The generalized Friedmann equation (\ref{EquationOfMotion2}) takes the
familiar coordinate form \cite{Hellaby:2002nx}%
\begin{equation}
\left( \dot{S}\right) ^{2}=\frac{2\mathcal{M}_{1}}{S}+f  \label{Friedmann3}
\end{equation}%
for all topologies of the 2d screen space. As a result, the function $f(r)$
represents an effective curvature term and can be also interpreted as twice
the energy per unit mass of the particles in the shells of matter at
constant $r$.

We conclude this section by noticing that although the Szekeres family of
models does not have, in general, Killing Vector Fields (KVFs), however does
admit a 3d set of \emph{intrinsic symmetries} (in the spirit e.g. of \cite%
{Collins-Szafron}). As we have seen in the previous section, the projection
tensor $p_{ab}$ corresponds to the metric of $\mathcal{X}$ with associated
covariant derivative \textquotedblleft $\bar{\nabla}_{a}$\textquotedblright
. For the quasi-spherical class, the screen space has constant and positive
curvature hence there exists a $SO(2)$ group of \emph{Intrinsic KVFs}
(IKVFs). Using the coordinate form (\ref{SzekeresMetricHellabyKrasinskiForm1}) of
the spacetime metric, we can verify that the vector fields 
\begin{equation}
\mathbf{X}_{1}=\left[ z-Z(r)\right] \partial _{y}-\left[ y-Y(r)\right]
\partial _{z}  \label{KVF1}
\end{equation}%
\begin{eqnarray}
\mathbf{X}_{2} &=&\left\{ 1+V^{-2}\left[ y-Y(r)\right] ^{2}-V^{-2}\left[
z-Z(r)\right] ^{2}\right\} \partial _{y}+  \nonumber \\
&&+2/V^{2}\left[ y-Y(r)\right] \left[ z-Z(r)\right] \partial _{z}
\label{KVF2}
\end{eqnarray}%
\begin{eqnarray}
\mathbf{X}_{3} &=&2/V^{2}\left[ y-Y(r)\right] \left[ z-Z(r)\right] \partial
_{y}+  \nonumber \\
&&+\left\{ 1+V^{-2}\left[ z-Z(r)\right] ^{2}-V^{-2}\left[ y-Y(r)\right]
^{2}\right\} \partial _{y}  \label{KVF3}
\end{eqnarray}%
satisfy $\bar{\nabla}_{(b}X_{\left( \alpha \right) a)}=0$ where $\alpha
=1,2,3$. We note that the \textquotedblleft isotropy\textquotedblright\ of
the quantities $w_{1}$, $w_{2}$ and $\mathcal{E}_{x}$ is the direct
consequence of the existence of the IKVFs.

\section{Discussion}

\setcounter{equation}{0}

Inhomogeneous models can be seen as exact \textquotedblleft
perturbations\textquotedblright , to first or higher order, of the standard
cosmological model therefore the analysis of their geometry and dynamics
will enlight many open questions regarding the effect of density
fluctuations in the evolution of the universe. With respect to FLRW model,
their richer variety of matter density profiles that we can accommodate,
they could also impose further restrictions leading to a better fine tunning
of the observational parameters \cite{Bolejko:2010ec}.

Towards to this goal, a covariant description of the significant class of
inhomogeneous Szekeres solutions has been presented using the 1+1+2
splitting of the spacetime. We have shown how the geometry and the dynamics
of the Szekeres models are described in terms of the kinematical quantities
of the spacelike congruences. It was possible to identify an \emph{effective
gravitational mass} constituted of the matter density of the dust fluid and
the contribution of the free gravitational field represented by the electric
part of the Weyl tensor.

We note that several issues regarding the topology of the Szekeres models
can also be successively dealed using the results of the present paper. For
example the existence of Apparent Horizons (AH) \cite{Hawking-EllisBook} and
Absolute Apparent Horizons (AAH) \cite{Krasinski:2012hv} has been considered
within the Szekeres models and their differences are illustrated. In the
case of the former and for the quasi-spherical subclass, an AH is located at 
$S=2M$. Indeed let us consider the non-geodesic\footnote{%
We easily verify using the covariant formalism of the present paper the most
general Szekeres model does not permit radial null geodesics \cite%
{Nolan-Debnath} which is a reminiscence of their non-symmetric structure.}
null radial vector field $n^{a}=u^{a}-x^{a}$ which represents the direction
of outgoing null curves. Expressing their distortion in terms of the $%
u^{a},x^{a}-$variables for an observer in the screen space we get 
\begin{equation}
p_{a}^{\hspace{0.2cm}i}p_{b}^{\hspace{0.2cm}j}n_{i;j}=p_{a}^{\hspace{0.2cm}%
i}p_{b}^{\hspace{0.2cm}j}\sigma _{ij}+Hp_{ab}-\frac{\mathcal{E}_{x}}{2}%
p_{ab}(\mathbf{x}).  \label{RadialNullVectorDerivatives1}
\end{equation}%
The null rays converge to a single curve when the induced expansion rate of
their surface area vanishes which implies that 
\begin{equation}
p^{ab}n_{a;b}=0\Rightarrow \sigma _{3}+H-\frac{\mathcal{E}_{x}}{2}%
=0\Rightarrow \left( \sigma _{3}+H\right) ^{2}=\frac{\mathcal{E}_{x}^{2}}{4}.
\label{AHCovariantCondition}
\end{equation}%
From equation (\ref{EquationOfMotion2}) or equivalently (\ref%
{2ScalarCurvature1}) we deduce that the surface $2\mathcal{M}_{1}=S$
corresponds to the AH of the Szekeres models. On the other hand for an AAH
the null vector $n^{a}=\sqrt{3}u^{a}\pm x^{a}+y^{a}+z^{a}$ is
\textquotedblleft almost\textquotedblright\ radial in the sense of \cite%
{Krasinski:2012hv} and in our approach must satisfy $p^{ab}n_{a;b}=0$.

On the other hand in order to explore the possibility of having accelerating
expansion in the SIIS\ models, we cannot use the standard definition for the
expansion $H$ because, the cosmological expansion in inhomogeneous
cosmologies depends on the \emph{direction} of observation and the Hubble
parameter $H$ corresponds to an average of the expansion rates and the
variation in different directions is hidden. Therefore, it seems more
natural to study and analyze the influence of the gravitational field on the
light beams that come from a \emph{specific direction} and passing through
certain fluid configurations. In this spirit let us, briefly, discuss how
the suggested program of the 1+1+2 splitting of the Szekeres models can be
conveniently applied for the determination of the luminosity distance $d_{L}$
of an isotropic source (S) 
\begin{equation}
d_{L}=\left( 1+z\right) ^{2}\frac{\sqrt{A_{0}}}{\sqrt{\Omega _{s}}}
\label{LuminosityDistance}
\end{equation}%
where $A_{0}$ is the physical cross-section of the light beam and $\Omega
_{s}$ is the solid angle formed from the source to the position of the
observer (O). The above expression allows the determination of the
luminosity distance $d_{L}$ as a function of the redshift $z$ through the
knowledge of $A_{0}$. The latter can be determined by using the Sachs
optical equations \cite{sachs}. These describe the evolution of the
expansion and shear of the beam along its null trajectory in a similar way
as in the case of the timelike and spacelike congruences. Then we can use
them in order to study light propagation in the SIIS models and
qualitatively estimate the induced increase of the luminosity distance
relative to the FLRW background thus generalizing the corresponding results
for the LT models.

In general geometries, the observed cosmological redshift $z$ is covariantly
defined by the differential relation \cite{Sachs1} 
\begin{equation}
d\ln (1+z)=-d\ln (k^{a}u_{a})  \label{Redshift}
\end{equation}%
where $k^{a}(\upsilon )$ is a \emph{null }and\emph{\ geodesic} vector
tangent to the congruence of null curves $\xi ^{a}(\upsilon )$ with affine
parameter $\upsilon $, representing the paths of the light rays originating
from S. Here $J_{a}$ is a \emph{spacelike} vector field pointing in the
direction of the observed light beam

\begin{equation}
k^{a}\equiv \frac{d\xi ^{a}}{d\upsilon }=\left[ k^{0}u^{a}+J^{a}\right] ,%
\hspace{0.5cm}k^{0}\equiv -k^{a}u_{a}  \label{NullVectorDefinition1}
\end{equation}

\begin{equation}
k^{a}k_{a}=0,\hspace{0.5cm}k_{a;b}k^{b}=0  \label{NullVectorDefinition2}
\end{equation}%
where the vector field $J^{a}$ is expressed as linear combination of the
eigen-basis $\{x^{a},y^{a},z^{a}\}$ 
\begin{equation}
J^{a}=\lambda _{1}x^{a}+\lambda _{2}y^{a}+\lambda _{3}z^{a}
\label{NullDirection}
\end{equation}%
and the dimensionless parameters $\lambda _{\alpha }$ satisfy the
orthonormality condition 
\begin{equation}
\sum\limits_{\alpha }\left( \lambda _{\alpha }\right) ^{2}=\left(
k^{0}\right) ^{2}.  \label{NullCondition1}
\end{equation}%
Essentially, the parameters $\lambda _{\alpha }$ correspond to the \emph{%
directional cosines} of the null geodesics \cite%
{Ellis-Nel-Sachs1-Maartens-Stoeger}\ relative to the orthonormal triad $%
\{x^{a},y^{a},z^{a}\}$ which is parallel-propagated \cite%
{Apostolopoulos:2006zn} along the world-line of the four-velocity $u^{a}$ of
the dust fluid. Then, the distortion of the light beam is encoded in the
shear, rotation rates and the expansion of the null geodesic congruence
which give rise to the average cross section $A$ and their evolution along $%
\xi ^{a}(\upsilon )$ can be conveniently formulated in terms of the
kinematical quantities of the timelike and spacelike congruences.

We conclude by noticing that in many situations with clear geometrical or
physical importance, apart from the existence of a preferred timelike
congruence, may also exist a preferred spacelike direction representing an
intrinsic geometrical/dynamical feature of a model or a physical situation.
Consequently, a further 1+2 splitting of the 3d space naturally arises,
leading to the concept of the 1+1+2 decomposition of the spacetime manifold.
As a result the 1+1+2 covariant analysis reported in this paper, revealed a
number of \textquotedblleft hidden\textquotedblright\ properties of the
Szekeres models with sound geometrical and physical usefulness and provide
an appropriate framework to study the effect of small or large
inhomogeneities in the cosmological expansion. This will be the subject of a
forthcoming work.\vskip0.5cm

\noindent \textbf{Acknowledgements}

\noindent The author would like to thank the anonymous reviewers for their
careful reading of the paper and their insightful comments and suggestions.

\end{document}